\documentclass[preprint,12pt]{elsarticle}

\usepackage[left=1in, right=1in, top=1in, bottom=1in]{geometry}
\usepackage{xcolor}
\usepackage{graphicx}
\usepackage{caption,subcaption}
\usepackage{amsmath,amsfonts,amssymb}
\usepackage{bm}
\usepackage{enumitem}
\usepackage{nomencl}
\usepackage{hyperref}

\hypersetup{
    colorlinks=true,
    linkcolor=blue,
    linktocpage=true}

\makenomenclature

\newcommand{\boldface}[1]{\boldsymbol{#1}}  

\newcommand{\bfk}{\boldface{k}}

\newcommand{\bfu}{\boldface{u}}
\newcommand{\bfv}{\boldface{v}}
\newcommand{\bfw}{\boldface{w}}
\newcommand{\bfx}{\boldface{x}}

\newcommand{\bfA}{\boldface{A}}

\newcommand{\bfC}{\boldface{C}}

\newcommand{\bfF}{\boldface{F}}

\newcommand{\bfQ}{\boldface{Q}}
\newcommand{\bfR}{\boldface{R}}

\newcommand{\calB}{\mathcal{B}}

\newcommand{\ti}{\text{i}}

\newcommand{\Rset}{\ensuremath{\mathbb{R}}}

\newlength{\boxwidth}
\def\dd{\;\!\mathrm{d}}

\def\btheorem{\begin{theorem}}
\def\etheorem{\end{theorem}}
\def\blemma{\begin{lemma}}
\def\elemma{\end{lemma}}
\def\bproposition{\begin{proposition}}
\def\eproposition{\end{proposition}}
\def\bcorollary{\begin{corollary}}
\def\ecorollary{\end{corollary}}
\def\bdefinition{\begin{definition}}
\def\edefinition{\end{definition}}
\def\bexample{\begin{example}}
\def\eexample{\end{example}}
\def\bremark{\begin{remark}}
\def\eremark{\end{remark}}

\newcommand{\ba}{\begin{array}}
\newcommand{\ea}{\end{array}}

\newcommand{\be}{\begin{equation}\nonumber}
\newcommand{\ee}{\end{equation}}
\newcommand{\beq}{\begin{eqnarray}}
\newcommand{\eeq}{\end{eqnarray}}
\newcommand{\bem}{\begin{multline}}
\newcommand{\eem}{\end{multline}}

\begin{document}

\title{Conformally Graded Metamaterials for Elastic Wave Guidance}

\author{Charles Dorn$^{1}$, Dennis M. Kochmann$^{{1},*}$}

\address{$^{1}$Mechanics \& Materials Lab, Department of Mechanical and Process Engineering, ETH Zürich, 8092 Zürich, Switzerland\\
$^*$Corresponding author, email: dmk@ethz.ch}

\begin{abstract}

Although metamaterials have been widely used for controlling elastic waves through bandgap engineering, the directed guidance of stress waves in non-periodic structures has remained a challenge. This work demonstrates that spatially graded metamaterials based on conformal mappings present a rich design space for controlling and attenuating wave motion –- without the need for bandgaps. Conformal mappings transform an elementary unit cell by scaling and rotation into graded lattices with approximately geometrically similar unit cells. This self-similarity allows for control over the local wave dispersion throughout the metamaterial. As a key mechanism, it is shown that elastic waves cannot propagate through graded unit cells with significant size differences, except at low frequencies. This is exploited to create low-pass elastic wave guides, extending beyond classical bandgap engineering, since bandgaps are not required to achieve wave guiding and attenuation. Experiments confirm the low-pass elastic wave filtering capability of a planar truss metamaterial with conformal grading. Finally, a systematic design of curved metamaterial surfaces is presented, providing a flexible framework for programming low-pass attenuation and wave guiding in three dimensions.

\end{abstract}

\maketitle

\newpage


\section{Introduction}

Lattices appear naturally in atomic crystals, where wave dispersion has been studied for decades \cite{brillouin1953wave} and plays a fundamental role in lattice vibrations, electrical and thermal conduction, and the refraction of light. Metamaterials do what nature cannot do: the careful design of lattices, which enables as-designed effective properties. Dispersive phenomena observed at the atomic scales, such as bandgaps (frequency bands where wave propagation is forbidden), can be achieved and exploited by metamaterials on a wide range of length scales. The design of periodic mechanical metamaterials, composed of repeating arrangements of beams, plates, shells, or composite materials, has been well-studied due to their appealing elastic wave attenuation capability arising from bandgaps \cite{hussein2014dynamics}. However, graded metamaterials with smooth spatial variation of unit cells have remained largely unexplored in the context of elastic wave propagation, despite the immensely increased design space compared to periodic architectures \cite{dorn2022ray}.

Motivated by the opportunity to leverage bandgaps for vibration suppression and wave guiding, significant efforts have been made to design and optimize periodic metamaterials with wide bandgaps that span desired frequencies \cite{sigmund2003systematic,halkjaer2006maximizing,vatanabe2014maximizing,bilal2011ultrawide,jensen2006maximal,dong2014topological}. By contrast, spatially graded architectures are less explored but offer a considerably larger design space, since both the unit cell content and its spatial variation can be controlled. For example, bandgaps of different unit cells in a graded metamaterial can combine to produce a wider effective bandgap. This phenomenon, sometimes referred to as rainbow trapping, has been demonstrated for elastic waves in one-dimensional \cite{chaplain2020delineating,chaplain2020topological,de2020graded,alshaqaq2020graded,alshaqaq2022programmable,zhao2022graded,rosafalco2023optimised} and two-dimensional \cite{trainiti2016wave,trainiti2018optical,aguzzi2022octet} metamaterials (where wave attenuation has been demonstrated along a specific direction only). While existing research highlights the promise of graded metamaterials for wave manipulation, systematic methods that explore the vast design space of spatially graded lattices have remained an open challenge.

In this work, we present metamaterials graded by conformal mappings, which we show to be effective for mechanical wave attenuation and wave guiding. Conformal mappings consist of a locally uniform scaling and rotation, ensuring that each unit cell in the graded lattice is approximately scaled and rotated. Hence, all unit cells are approximately geometrically similar. Due to this self-similarity, we show that waves cannot propagate along trajectories that see a significant change in unit cell size, except at low frequencies. A simple yet powerful design principle follows: designing regions with a significant unit cell size difference ensures that waves cannot propagate between these regions. As a result, certain regions can be isolated from high-frequency wave propagation. Additionally, we show that attenuation via spatial grading is achievable in the absence of bandgaps, as long as neighboring dispersion surfaces are non-intersecting.

Our use of conformal mappings for mechanical metamaterials is distinct from their extensive use for optical metamaterials. Transformation optics often relies on conformal mappings to achieve practical effective material parameters \cite{leonhardt2006optical,vasic2010controlling,yao2011designing,xu2015conformal} (similarly for transformation acoustics \cite{dong2017realization,sun2019quasi,dong2020bioinspired,yu2020active}). However, transformation optics and acoustics typically apply in the non-dispersive setting \cite{mccall2018roadmap}, requiring sub-wavelength unit cells. Furthermore, they do not directly generalize to elastodynamics due to a lack of form invariance \cite{milton2006cloaking}. 

In the elastodynamic setting, conformal maps have primarily been utilized in the low-frequency and long-wavelength (with respect to a unit cell) regime by relying on effective continuum properties \cite{chang2012elastic,chen2016design,gao2019manipulating,nassar2019isotropic,nassar2020polar}, thus missing out on the rich dispersive behavior arising in the short-wavelength regime that is key to mechanical metamaterials. In contrast to previous work, we utilize conformal mappings to control waves with short wavelengths. To this end, with inspiration from the discrete geometry literature \cite{springborn2008conformal,crane2011spin,Crane:2020:DCG}, we design planar and curved conformally graded metamaterial surfaces. While conformal mappings represent only a fraction of the spatial grading design space, we show that their wave dynamics can be engineered using a simple but general design procedure.

The resulting spatially graded metamaterials appeal to a wide range of applications across scales. Their broad effective bandgaps are practical for vibration suppression, e.g., to shield infrastructure from seismic waves \cite{colombi2016seismic}. Grading is also appealing for energy harvesting by trapping or focusing waves \cite{tol2017phononic,de2021graded}. Additionally, frequency-selective sensing and signal processing capabilities are enabled by spatial grading \cite{zhao2019compact,chen2021tunable}. The conformal architectures presented in this work enhance the applicability of graded metamaterials by offering low-pass attenuation capabilities without relying on bandgaps. Furthermore, we propose a systematic design method for curved metamaterial surfaces, which offers a flexible tool for programmable wave guidance and attenuation.


\begin{figure*}[t]
    \centering
    \includegraphics[width=\linewidth]{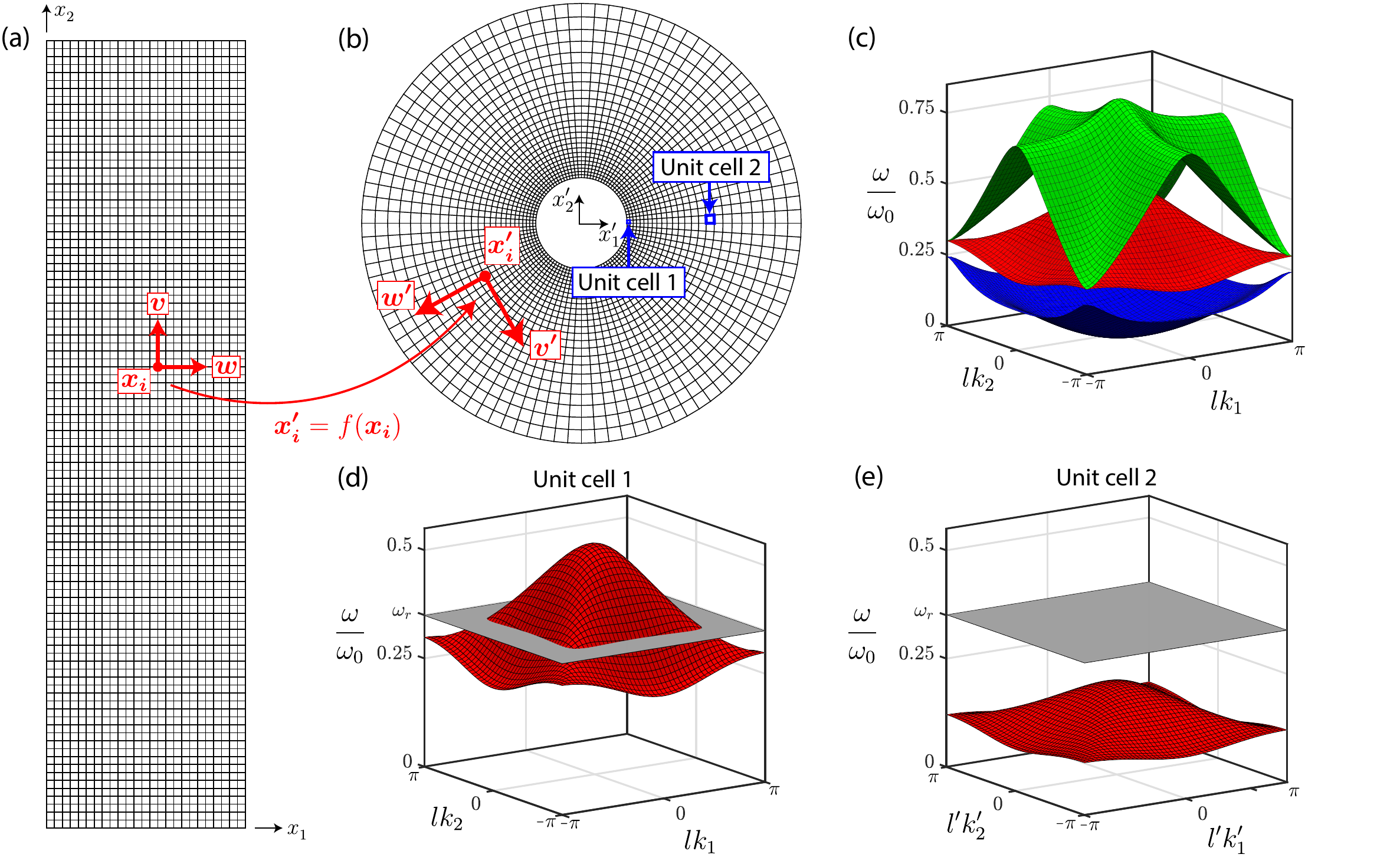}
    \caption{Planar truss lattice for low-pass attenuation: (a) Reference periodic truss lattice. (b) Graded truss lattice corresponding to a conformal mapping of the reference lattice. (c) Dispersion relations of the reference lattice with unit cell side length $l$ and $\omega_0=l^{-1}\sqrt{E/\rho}$. (d) Second dispersion surface of unit cell~1, and (e) of unit cell~2 highlighted in (b), with the $\omega_r$ level set highlighted.}
    \label{fig:figure1}
\end{figure*}

\section{Conformally Graded Metamaterials}

We design spatially graded lattices through a mapping $f:\mathcal{M}\mapsto\mathbb{R}^3$, which transforms a planar, \textit{periodic} reference lattice $\mathcal{M}\subset\Rset^2$ into a \textit{graded} lattice $\mathcal{M}'\subset \mathbb{R}^3$. Surface $\mathcal{M}'$ can be flat or curved (both cases will be illustrated). We denote points in the reference lattice by $\bfx \in \mathcal{M}$ and their mapped position in the graded lattice by $\bfx'=f(\bfx)\in\mathcal{M}'$.

The differential $\dd f = \frac{\dd \bfx'}{\dd \bfx}$ of the mapping defines how a vector $\bfv\in\mathbb{R}^2$ maps to the vector $\bfv'=\dd f(\bfv)\in\mathbb{R}^3$ (i.e., it characterizes how vectors tangent to the reference lattice map onto the corresponding vectors tangent to the graded lattice). The mapping is \textit{conformal}, if $\dd f(\bfv) \cdot \dd f(\bfw)=J^2 \bfv \cdot \bfw$, where $J>0$ is the \textit{conformal scaling factor}. 

Conformal mappings are chosen specifically, because they consist solely of a local uniform scaling (since for any vector $\bfv$ we have $|\dd f(\bfv)|=J|\bfv|$) and a rotation (since the angle between vectors is unchanged by the mapping) of the unit cells. Fig.~\ref{fig:figure1}ab illustrates a conformal mapping in the case where $\mathcal{M}'$ is planar. Here, the vectors $\bfv$ and $ \bfw$, which are attached to the reference lattice at point $\bfx_i$, are uniformly scaled and rotated by the mapping to vectors $\bfv'$ and $\bfw'$, while the angle between them is preserved. For a more extensive treatment of conformal mappings, see~\cite{nehari2012conformal,gu2008computational}.

We assume that in the transformed lattice the unit cells vary slowly in space, so that $\dd f$ can be taken as constant throughout any given unit cell. Under this assumption, each unit cell undergoes only uniform scaling by a factor $J$ and a rotation. This assumption allows the graded lattice to be modeled as \textit{locally periodic}, so that wave propagation is governed by the \textit{local dispersion relations} \cite{dorn2022ray}. That is, the local dispersion relations at a given unit cell on the graded lattice are calculated assuming periodicity of that unit cell, which is a valid approximation for slow spatial gradings.

Conformal grading yields a special case of local periodicity since, under the above assumptions, all unit cells are geometrically similar to the unit cell of the reference lattice. Consider a reference unit cell with dispersion relations $\omega(\bfk)$, relating the wave vector $\bfk \in \mathbb{R}^2$ to the frequency $\omega$ of an elastic plane wave. A unit cell in the transformed lattice has local dispersion relations $\omega'(\bfk')$. Since this unit cell is a scaled and rotated copy of the reference unit cell, its dispersion relations take the form
\begin{equation} \label{eq:dispersion_scaling}
    \omega' (\bfk') = J^{-1} \omega (\bfk), 
\end{equation}
where $\bfk'$ is the wave vector in the transformed coordinates (a derivation is provided in Section~1 of the Supplementary Material, which holds generally for linear elastic waves). For the example mapping in Fig.~1, the dispersion relations of the reference lattice (Fig.~\ref{fig:figure1}c) are hence simply re-scaled to determine the local dispersion relations of any unit cell in the graded lattice, including unit cells~1 (Fig.~\ref{fig:figure1}d) and 2 (Fig.~\ref{fig:figure1}e).

To understand how to exploit this inverse frequency scaling relation, we view wave propagation from the perspective of ray theory. Ray theory has long been used to model wave propagation in many fields, ranging from optics \cite{born2013principles} to seismology \cite{cerveny2001seismic}, where it provides approximate solutions to a wave equation along characteristic ray trajectories. We recently extended it to graded elastic metamaterials \cite{dorn2022ray}, under the assumption that wavelengths are significantly smaller than the length scale associated with the spatial grading of the unit cells. Ray theory is valid in the context of high frequencies and short wavelengths, i.e., when unit cells change slowly in space relative to the wavelength, which is the context considered here.

We rely on two key properties of rays to reveal the special properties of conformally graded lattices in the linear elastic regime: 
\begin{enumerate}[label=(\roman*)]
\item A ray propagates at a fixed frequency.
\item A ray is associated with a particular dispersion surface (i.e., a particular mode). It cannot switch between modes as long as the unit cells vary slowly in space and no frequency degeneracies are encountered. 
\end{enumerate}
Property (i) holds as long as the local dispersion relations are time-invariant \cite{thorne2017modern}, which is generally the case for elastic media. Property (ii) is a re-statement of the adiabatic theorem, which is well-known in quantum mechanics \cite{sakurai1995modern} but also applies to elastodynamics \cite{nassar2018quantization}.

Combining Properties (i) and (ii) with the dispersion scaling law of Eq.~\eqref{eq:dispersion_scaling} admits a general observation about elastic wave propagation in conformally graded lattices: \textit{high-frequency waves cannot propagate along trajectories that see significant changes in unit cell size}. This becomes evident as follows. By Eq.~\eqref{eq:dispersion_scaling}, local dispersion surfaces everywhere are a scaled copy of the reference dispersion surfaces. By Properties (i) and (ii), a ray is confined to a level set of a given mode's dispersion surface corresponding to the ray's fixed frequency $\omega_r$ (e.g., in Fig.~\ref{fig:figure1}d, a ray's wave vector must lie on the intersection of the dispersion surface and the ray's fixed frequency $\omega_r$). If that dispersion surface has a nonzero minimum frequency, it is always possible to scale that surface up or down such that it no longer intersects $\omega_r$ (e.g., in Fig.~\ref{fig:figure1}e there is no intersection between $\omega_r$ and the scaled dispersion surface). This scaling argument, however, does not apply at low frequencies of dispersion surfaces emerging from $\omega=0$ (the so-called acoustic modes).

By this scaling argument, attenuation via spatial grading is achievable when adjacent dispersion surfaces are non-intersecting. This includes bandgaps, which ensure that adjacent dispersion surfaces do not intersect. However, attenuation is also possible in the absence of a bandgap, if adjacent dispersion surfaces are non-intersecting, as we show in the following examples.


\section{A Planar Truss Lattice for Low-Pass Attenuation}

An instructive example that demonstrates the use of a conformal grading for wave attenuation is a planar truss lattice that acts as a low-pass filter of elastic waves. Consider the conformal mapping
\begin{equation} \label{eq:exp_map}
    z' = e^z,
\end{equation}
where $z'=x_1'+\ti x_2'$, $z=x_1+\ti x_2$ in 2D, and $\ti=\sqrt{-1}$. Fig.~\ref{fig:figure1} illustrates this mapping, which conformally transforms the reference lattice with square unit cells (Fig.~\ref{fig:figure1}a) to a radial lattice (Fig.~\ref{fig:figure1}b). In the transformed lattice, the outermost unit cells are $J_\text{o}=3.51$ times larger than the innermost ones, which have $J_\text{i}=1$, where $J_\text{o}$ and $J_\text{i}$ are scaling factors with respect to the reference lattice.

\begin{figure*}[t!]
    \centering
    \includegraphics[width=\linewidth]{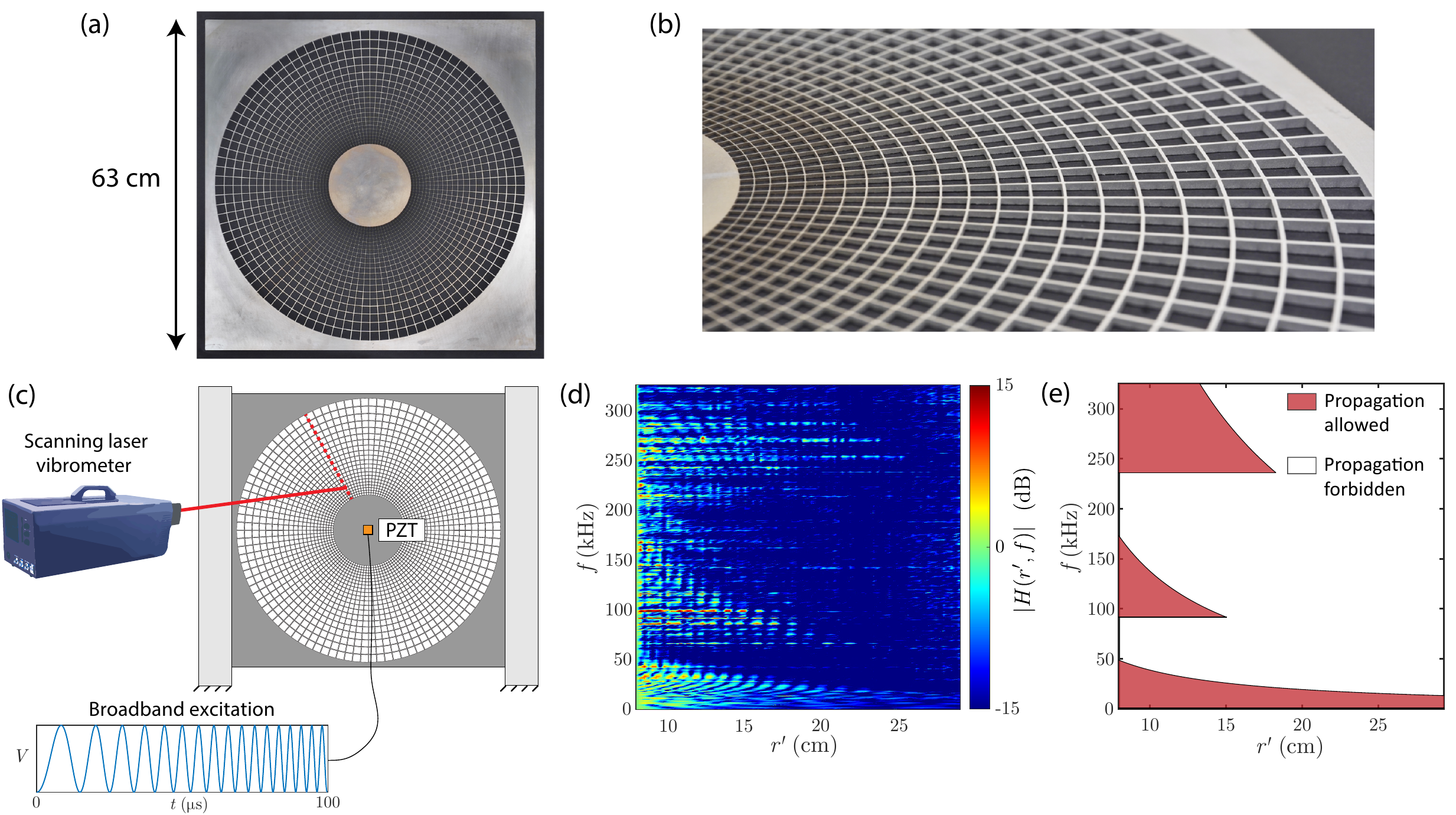}
    \caption{Experimental confirmation of the low-pass attenuation in a conformally graded truss: truss lattice prototype from the top view (a) and a close-up view (b). (c) Experimental setup: a piezoelectric transducer excites the lattice, whose response is measured by a scanning laser Doppler vibrometer along radial lines. (d) Measured FRF amplitude along a radial line. (e) Predicted regions of allowed and forbidden propagation based on dispersion relation scaling.}
    \label{fig:figure2}
\end{figure*}

Each unit cell consists of slender linear elastic beams along its four edges. Dispersion relations for the reference unit cell are computed numerically using a beam finite element model of a unit cell (Supplementary Material, Section~2). The first three dispersion surfaces corresponding to out-of-plane modes are plotted in Fig.~\ref{fig:figure1}c over the first Brillouin zone, with frequencies normalized by $\omega_0=l^{-1}\sqrt{E/\rho}$, where $l$ is the edge length of the reference unit cell, and $E$ and $\rho$ are Young's modulus and the mass density of the beams' base material. The lowest modes of truss lattices are decoupled into in- and out-of-plane modes \cite{zelhofer2017acoustic}; by considering out-of-plane excitation, only the out-of-plane modes must be considered. 

This conformally graded truss lattice acts as a low-pass filter for waves applied to the center of the lattice; waves above a cutoff frequency cannot propagate from the inner to the outer radius. To illustrate this, consider a mode~2 ray that starts in unit cell~1 of Fig.~\ref{fig:figure1} and propagates radially outward at the excitation frequency $\omega_r$. In unit cell~1, the wave vector of this ray must lie on the intersection of the mode~2 (red) dispersion surface and the $\omega_r$ level-set (gray), as depicted in Fig.~\ref{fig:figure1}d. Moving to unit cell~2, which is 2.73 times larger than unit cell~1, its dispersion surface is obtained by scaling the frequency of unit cell~1's dispersion surface by a factor of $1/2.73$. At unit cell~2, there is no intersection between the mode 2's dispersion surface and the $\omega_r$ level set (Fig.~\ref{fig:figure1}e). Therefore, a mode~2 ray starting at unit cell~1 with frequency $\omega_r$ is forbidden in unit cell~2. Supplementary Video~S1 visualizes the dispersion surface as the unit cell is re-scaled, highlighting the critical scaling factor where propagation at frequency $\omega_r$ becomes forbidden.

By applying this scaling argument more generally, rays of all frequencies (except those approaching $\omega=0$ of mode 1) become forbidden for a sufficiently large scaling factor, corresponding to when the dispersion surface is scaled to no longer intersect the frequency of the ray. Thus, rays starting at unit cell~1 will reach some maximum radius, beyond which propagation is forbidden. Based on this dispersion scaling argument, the critical radius is computed across a range of frequencies and plotted in Fig.~\ref{fig:figure2}e. Regions where propagation is allowed are highlighted in red, and regions where a wave starting at the inner radius is forbidden are white. Details of the scaling analysis are given in Section~2.3 of the Supplementary Material, where care is taken to account for mode conversion due to frequency degeneracies where dispersion surfaces intersect. 

We experimentally demonstrate the attenuation capability of this geometry, using the prototype in Fig.~\ref{fig:figure2}a, which is made of aluminum and has a tapered thickness to ensure that all unit cells are geometrically similar (see the close-up view in Fig.~\ref{fig:figure2}b). Broadband excitation was applied to the center by a piezoelectric (PZT) transducer in the out-of-plane direction. The out-of-plane displacement was measured with a scanning laser Doppler vibrometer, as illustrated in Fig.~\ref{fig:figure2}c. Details of the fabrication and experiment are presented in Section~2.2 of the Supplementary Material. 

The experimental results in Fig.~\ref{fig:figure2}d show the amplitude of the frequency response function (FRF) \cite{ewins2009modal}, denoted $H(r',f)$, between the displacement at radial coordinate $r'$ and the inner radius. This provides a measure of attenuation as a wave travels outward from the inner radius by interpreting the displacement signal at the inner radius as the input to the truss. Regions with high FRF amplitude in Fig.~\ref{fig:figure2}d agree closely with allowable propagation regions in Fig.~\ref{fig:figure2}e, whereas regions of low values correspond to the predicted regions of forbidden propagation. Based on the dispersion relation scaling analysis, the predicted low-pass cutoff frequency is 14.7~kHz; higher frequencies are predicted to be attenuated before reaching the outer radius of the lattice. This prediction agrees with the experimental low-pass cutoff frequency, which is observed to be about 16~kHz.

\begin{figure*}[!b]
    \includegraphics[width=\linewidth]{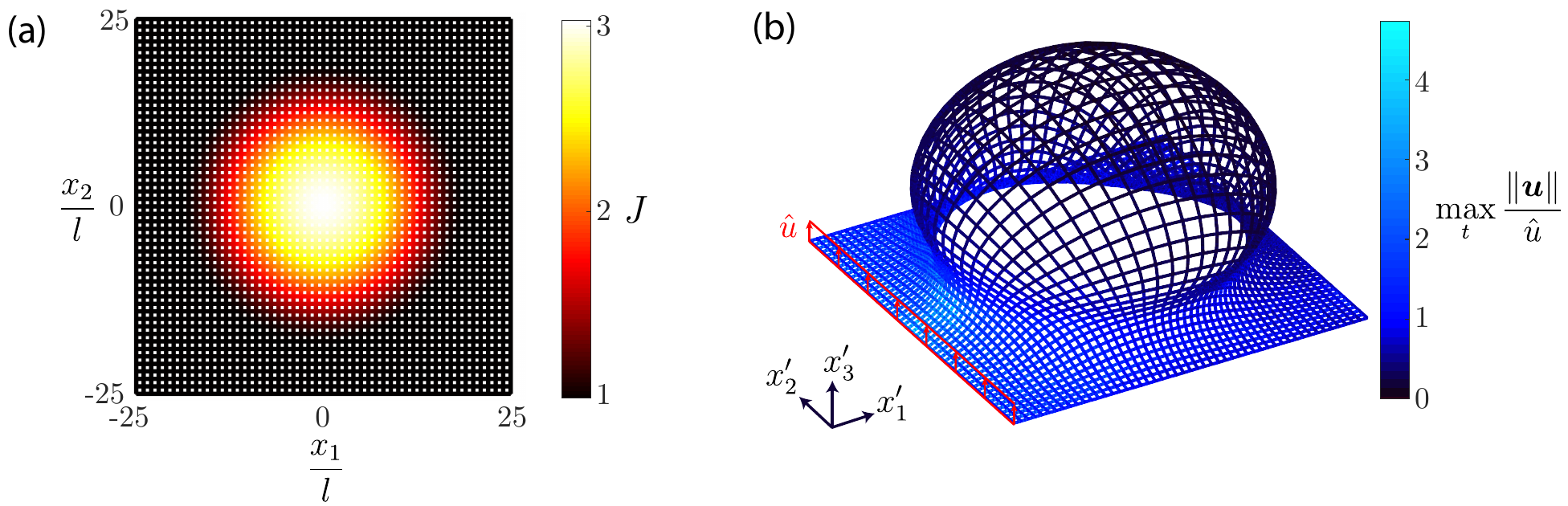}
    \caption{Conformal bump example. (a) Target scaling factor distribution plotted on the reference lattice. (b) Optimal conformal truss lattice realizing the prescribed scaling. Color indicates the maximum amplitude during a finite element simulation with broadband excitation applied on the lower left edge (highlighted in red) with amplitude $\hat u$.}
    \label{fig:figure3}
\end{figure*}


\section{Curved Metamaterial Surfaces}

In addition to the above plane-to-plane case, the same physical principles apply to conformal mappings that transform a planar surface into a curved surface, for which we propose a systematic design framework for curved conformal lattice surfaces. Based on the guiding principle that waves do not propagate between unit cells with a significant scaling difference, our objective is to control the conformal scaling factor distribution in the lattice in order to either isolate a region from incident high-frequency waves or to confine waves in a specified region. The advantage of curved surfaces is that curvature offers a richer design space than planar lattices, allowing for elaborate scaling factor distributions to be realized as follows. Based on a desired unit cell scaling distribution, a target conformal scaling factor distribution $J(\bfx)$ is first prescribed on a planar reference lattice. Next, an optimization problem identifies a transformed three-dimensional (3D) lattice that best realizes the prescribed scaling factor distribution.

We approach this problem from the perspective of discrete differential geometry, which naturally follows from the discrete nature of the lattice. In the discrete setting, the target scaling distribution is achieved by re-scaling the edges from the reference length $l$ to a target length~$\tilde{l}$. The target edge lengths are carefully chosen to satisfy discrete conformal equivalence \cite{springborn2008conformal}, as detailed in the Supplemental Material, Section~3.1.

The problem of computing a 3D embedding (i.e., vertex coordinates $\bfx'$) of a discrete surface given its discrete metric (i.e., edge lengths $\tilde{l}$) has been addressed in the discrete differential geometry literature \cite{isenburg2001connectivity,wang2012linear,boscaini2015shape,chern2018shape}. Inspired by these approaches, to compute the transformed vertex coordinates $\bfx'$ we pose the unconstrained minimization
\begin{equation} 
\min_{\bfx'}E(\bfx'),
\end{equation}
where
\begin{equation} \label{eq:optimization}
    E(\bfx')=\sum_m k \big(l_m(\bfx')-\tilde{l}_m\big)^2 + \sum_n \kappa\, \theta_n(\bfx')^2.
\end{equation}
Here, $m$ indexes all edges and $n$ indexes the interior edges, which are not on the boundary. This optimization can be interpreted as an energy minimization problem by imagining that an extensional spring is placed on each edge, with stiffness $k$ and rest length $\tilde{l}$. In practice, a regularization term is required to achieve smooth solutions, otherwise the optimal surfaces will be `crumpled' \cite{chern2018shape}. To this end, the second term in Eq.~\eqref{eq:optimization} effectively places a torsional spring on all interior edges with torsional stiffness $\kappa$ and zero rest angle (the latter corresponds to a flat configuration when the faces connected to the edge are co-planar). This term penalizes crumpled configurations in favor of smoother solutions. Smooth unit cell variations are further essential to avoid sharp kinks, which may invalidate Property~(ii) by leading to mode conversion or in-plane and out-of-plane mode coupling.

\begin{figure*}[!t]
    \includegraphics[width=\linewidth]{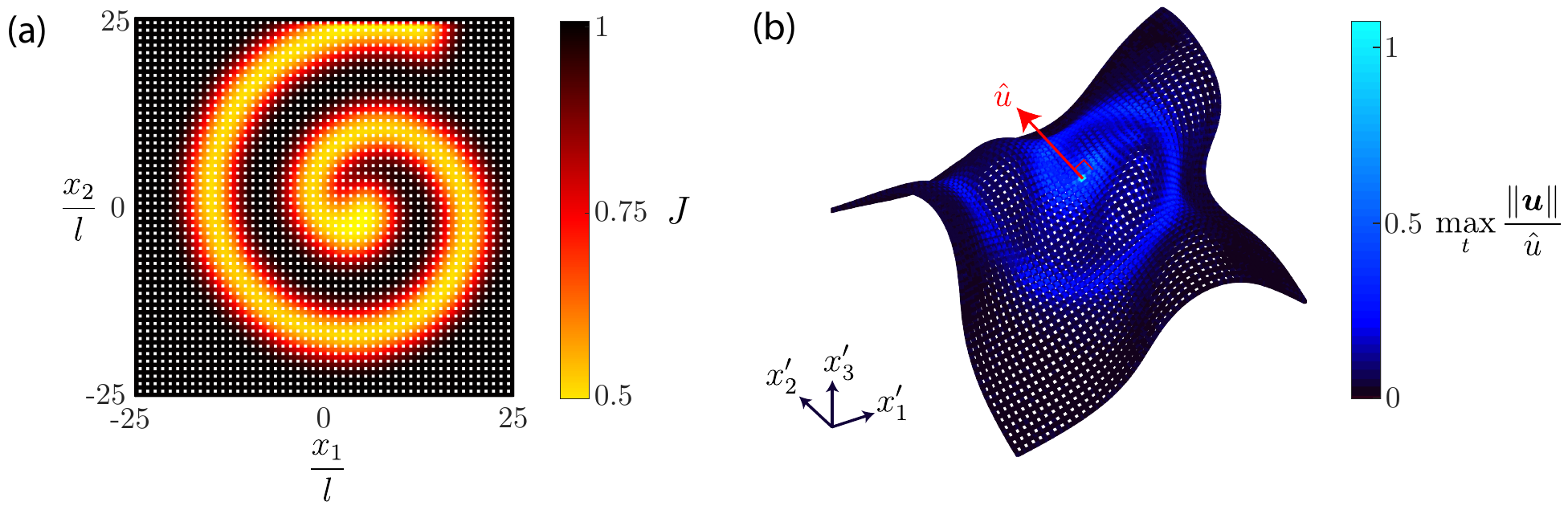}
    \caption{Conformal spiral example. (a) Target conformal scaling factor distribution plotted on the reference lattice. (b) Optimal conformal truss lattice realizing the prescribed scaling. Color indicates the maximum amplitude during a finite element simulation with a harmonic point excitation of displacement amplitude $\hat u$ perpendicular to the surface (along the red arrow).}
    \label{fig:figure4}
\end{figure*}

Fig.~\ref{fig:figure3} presents the example of a \textit{conformal bump}. The reference lattice is planar with uniform square unit cells comprised of beams along each edge. The scaling factor is defined according to a radial bump function and is plotted in Fig.~\ref{fig:figure3}a on the reference lattice. It leads to large unit cells in the center with a maximum scaling factor of $J=3.0$, surrounded by smaller unit cells of $J=1$ outside of the bump; the optimal 3D geometry is plotted in Fig.~\ref{fig:figure3}b. (Constraints were added to fix the four boundaries of the lattice; for implementation details, see Section~3 of the Supplementary Material.)

The bump in the center of the transformed lattice, which contains large unit cells, is isolated from high-frequency waves initiated on the boundaries of the lattice, whose smaller unit cells support those frequencies. To demonstrate this effect, a transient finite element simulation was performed. In the simulation, broadband displacement excitation was applied on the lower left (red) edge in the $x_3'$-direction with magnitude $\hat{u}$. Beam elements were used to model the lattice; see the Supplementary Material, Section~3.4 for details. The suppression of waves in the bump region is evident from the color map of Fig.~\ref{fig:figure3}b, which shows the maximum displacement magnitude at each point over the duration of the simulation (see also Supplementary Video~S2). Importantly, this mechanism is not restricted to a narrow range of frequencies (as is often the case when leveraging bandgaps to manipulate waves). Instead \textit{all} frequencies above the critical frequency (which is controlled by the unit cell scaling factor distribution) are attenuated in the bump region. This effectively extends the benefits of rainbow trapping to, in principle, arbitrarily defined regions. 

This type of architecture is appealing for vibration isolation applications. For example, sensitive equipment could be mounted to the bump for shielding from high-frequency vibrations. To potentially achieve full attenuation across all frequencies, one could consider metamaterials connected to a foundation, which can exhibit bandgaps starting at zero frequency \cite{liu2012wave}, combined with conformal grading.

Beyond merely attenuating waves, conformal grading also provides a means of designing wave guides. A second example of a \textit{conformal spiral} is presented to demonstrate wave guiding. In this example, a region of small unit cell scaling in the shape of an Archimedean spiral is prescribed, which is plotted on the reference lattice in Fig.~\ref{fig:figure4}a. In the spiral region, the smallest scaling factor is $J=0.47$, while a factor of $J=1$ is prescribed outside of the spiral region. The optimal lattice surface is plotted in Fig.~\ref{fig:figure4}b, which takes the form of a wavy surface to accommodate the smaller unit cells along the spiral.

In this example, the path of small unit cells along the spiral is capable of confining waves. We demonstrate this with a transient finite element simulation, where a harmonic displacement point excitation of magnitude $\hat{u}$ is applied at the center of the spiral in the direction locally normal to the surface at that point. The frequency of the excitation is chosen to be near the top of the lowest dispersion surface, such that only a small increase in unit cell size forbids propagation. The resulting wave emanating from the point source follows the path of small unit cells and is thus guided along the spiral. The coloring of Fig.~\ref{fig:figure4}b shows the maximum displacement magnitude throughout the lattice during the simulation (see also Supplementary Video~S3). This example demonstrates that the spatially graded lattice can be leveraged to effectively steer waves by designing regions of forbidden propagation. 

While both examples confirm that waves are attenuated in regions with a large unit cell scaling factor compared to the excitation location, perfect attenuation is not achieved, as may be expected for a number of reasons. The assumption of local periodicity is an approximation, as is the assumption that the `locally flat' dispersion relations are valid for curved lattice surfaces. Finally, the unit cells are not exactly geometrically similar in these two examples. Nevertheless, our assumptions are sufficiently valid to produce the desired wave confinement and attenuation behavior.

\section{Conclusion}

We have shown that conformally graded metamaterials effectively attenuate mechanical waves above a controllable cut-off frequency and hence can be designed to behave as effective waveguides---without requiring bandgaps. While metamaterials with broadband attenuation capability have been pursued by a growing body of research, conformally graded architectures go beyond broadband to achieve low-pass attenuation. Furthermore, our dispersion scaling analysis is more general than tracking how bandgaps shift with spatial unit cell variations, which is typically done to study attenuation in graded architectures \cite{trainiti2016wave,trainiti2018optical,aguzzi2022octet}. Specifically, a bandgap between two dispersion surfaces is not required to achieve attenuation, but only the less strict requirement that there is no intersection between neighboring dispersion surfaces. For example, there is no bandgap between the lowest two dispersion surfaces of Fig.~\ref{fig:figure1}c; yet, since there is no intersection between them, attenuation via spatial grading is achieved.

While an experimental demonstration confirms the low-pass attenuation capability of a planar lattice, curved lattice surfaces offer a wide design space for conformal grading, whose systematic inverse design shows promise for vibration isolation applications across length scales. 
This work represents only a small step into the vast and largely unexplored design space of graded and curved metamaterial surfaces. The effectiveness of conformal grading suggests that more general types of grading are promising for linear elastic wave manipulation.

\section*{Acknowledgments}
This work was supported by an ETH Zürich Postdoctoral Fellowship. The authors thank Vignesh Kannan for assistance with the experimental setup and Bastian Telgen for assistance with dispersion relation computations.


\newpage

\setcounter{figure}{0}
\setcounter{equation}{0}

\renewcommand{\theequation}{S\arabic{equation}}
\renewcommand{\thefigure}{S\arabic{figure}}
\renewcommand{\thesection}{S\fpeval{\arabic{section}-5}}

\begin{center}
\large  Supplementary material for: ``\textbf{Conformally Graded Metamaterials for Elastic Wave Guidance}'' \\by Charles Dorn and Dennis M.~Kochmann\\
\end{center}
\normalsize


\section{Dispersion relation scaling derivation}

A key observation explored in this work is that geometrically similar unit cells have dispersion relations that scale inversely with the unit cell's size, which is stated in Eq.~(1) of the main text. Here, we provide a derivation of this result. To this end, we briefly review elastic wave propagation in periodic solids, before focusing on changes to the dispersion relations when geometrically transforming the unit cell.


\subsection{Waves in periodic solids}
Let $\mathcal{B}\subset\mathbb{R}^3$ be a linear elastic body with spatial coordinates $\bfx\in\calB$. The mass density $\rho(\bfx)$ and fourth-order elasticity tensor $\bfC(\bfx)$ are periodic with a unit cell $\Omega$. The governing elastodynamic equation of linear momentum balance is
\begin{align} \label{eq:elastodynamic}
    \rho \ddot{u}_i = \left(C_{ijkl} u_{k,l}\right)_{,j},
\end{align}
where $\bfu(\bfx,t):\mathcal{B}\times\mathbb{R} \rightarrow \mathbb{R}^3$ is the displacement field and $t$ is time. Here and in the following, we use Einstein's summation convention, and a comma index denotes differentiation; i.e., $(\cdot)_{,i} = \frac{\partial (\cdot)}{\partial x_i}$.

When using Bloch wave analysis, only a single unit cell needs to be considered. A displacement ansatz is made for one unit cell in the form of a plane wave
\begin{equation} \label{eq:ansatz}
    \bfu(\bfx,t) = \bfA e^{\text{i} \varphi},
\end{equation}
where $\varphi = \bfk\cdot \bfx - \omega t$ is the phase, $\bfA\in\mathbb{C}^3$ is the complex-valued amplitude vector, and $\text{i}=\sqrt{-1}$. Inserting this ansatz into Eq.~\eqref{eq:elastodynamic} and simplifying leads to the Bloch eigenvalue problem
\begin{align} \label{eq:Bloch_EVP_periodic}
    \boldsymbol{\mathcal{L}}^B [ \bfA ] + \omega^2 \rho \bfA = \boldsymbol{0},
\end{align}
where the \textit{Bloch operator} is defined as
\begin{align} \label{eq:Bloch_operator}
    \mathcal{L}^B_i [\bfA] =
    \frac{\partial C_{ijkl}}{\partial x_j} \left(  \frac{\partial}{\partial x_l} + \text{i} k_l \right) A_k + C_{ijkl} \left( \frac{\partial^2}{\partial x_l \partial x_j} + \text{i} k_j \frac{\partial}{\partial x_l} +  \text{i}  k_l \frac{\partial}{\partial x_j} - k_l k_j  \right) A_k.
\end{align}
A standard way to compute the dispersion relations is to convert Eq. \eqref{eq:Bloch_EVP_periodic} to the weak form, apply Bloch boundary conditions, and solve the eigenvalue problem numerically using, for example, a finite element model of a unit cell. This process is described, e.g., in \cite{hussein2014dynamics}.


\subsection{Dispersion relations of a transformed unit cell}
We proceed to derive how the dispersion relations are related between two unit cells that are geometrically similar. Specifically, we will show that the Bloch eigenvalue problem preserves its form under uniform scaling and rotation, but with the eigenvalues scaled inversely by the uniform scaling factor. Lattices that are periodic in two dimensions are considered, though we model the unit cells as three-dimensional as we consider out-of-plane displacements.

\begin{figure}
    \centering
    \includegraphics[width=\linewidth]{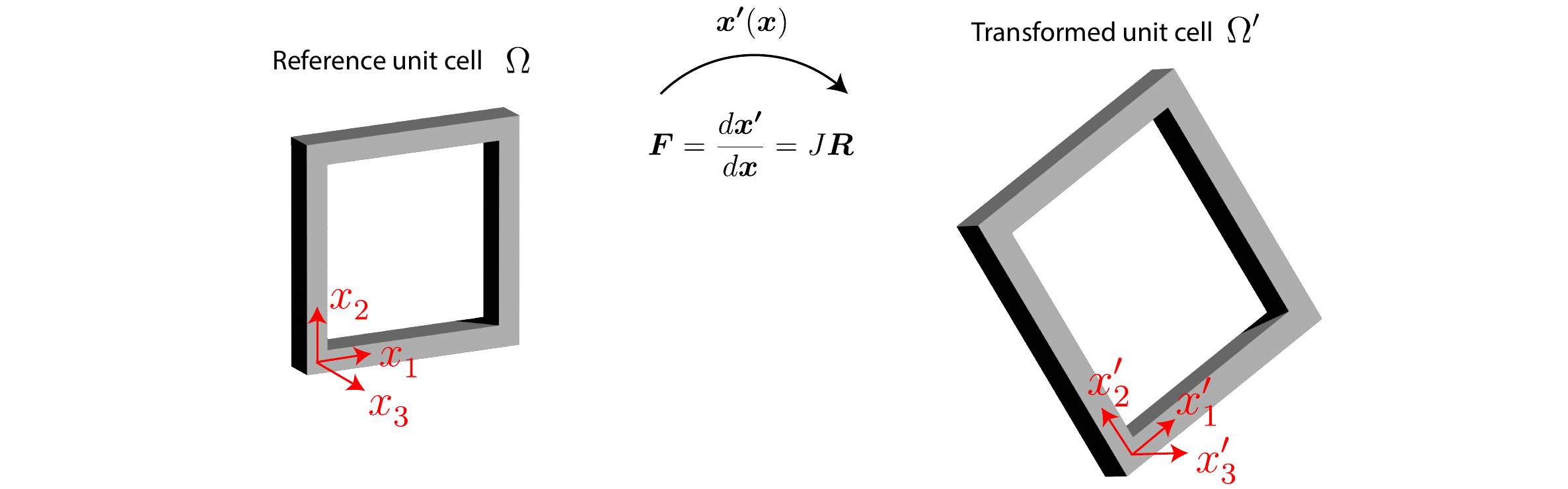}
    \caption{Reference unit cell $\Omega$ alongside the scaled and rotated unit cell $\Omega'$.}
    \label{fig:figureS1}
\end{figure}

The unit cell $\Omega$ is assumed periodic in coordinates $x_1$ and $x_2$, with $x_3$ representing the out-of-plane coordinate. A second unit cell $\Omega'$ is periodic in coordinates $x'_1$ and $x'_2$ with out-of-plane coordinate $x'_3$. Unit cell $\Omega'$ is geometrically similar to $\Omega$; its geometry is obtained by uniformly scaling and rotating $\Omega$. The transformation from $\Omega$ to $\Omega'$ can be expressed through a coordinate transformation $\bfx'=\bfx'(\bfx)$, whose gradient $\bfF\in\mathbb{R}^{3\times 3}$ is constant over the unit cell and consists of a scaling factor $J\in \mathbb{R}$ and a rotation matrix $\bfR\in\mathbb{R}^{3\times 3}$, such that
\begin{align} \label{eq:jacobian}
    \bfF = \frac{d\bfx'}{d\bfx}= J\bfR.
\end{align}
Fig.~\ref{fig:figureS1} shows an example of two such unit cells, for the case of a square unit cell with a beam along each edge.

Note that the conformal mappings discussed in the main text do not dictate how the thickness of the unit cell transforms. We choose to enforce that unit cell thickness also scales with $J$ to ensure geometric similarity of all unit cells. Consequently, all examples in this work involve graded lattices with variable thickness (i.e., both the in-plane dimensions and the out-of-plane thickness scale with $J$).

The rotation matrix $\bfR$ represents a rotation in the plane of the unit cell and has the form
\begin{align}
    \bfR = \begin{bmatrix} Q_{11} &  Q_{12} & 0 \\ Q_{21} &  Q_{22} & 0 \\ 0 & 0 & 1 \end{bmatrix},
\end{align}
where $\bfQ$ is a two-dimensional rotation matrix. 

Using the transformation between $\Omega$ and $\Omega'$, the Bloch eigenvalue problem for the reference unit cell can be rewritten in the transformed coordinates. Our aim is to show that this is identical to the Bloch eigenvalue problem written directly for the transformed unit cell but with frequency scaled by $J^{-1}$. First, we define how the spatial derivatives and wave vector transform. In terms of $\bfF$, we can relate spatial derivatives and wave vector in the reference domain to those in the transformed domain as
\begin{align} \label{eq:ddx_transform}
    \frac{\partial}{\partial x_i}& = \frac{\partial}{\partial x_{i'}'}\frac{\partial x_{i'}'}{\partial x_i} = \frac{\partial}{\partial x_{i'}'}F_{i'i}
\end{align}
and
\begin{align} \label{eq:k_transform}
    k_i &= \frac{\partial \varphi}{\partial x_i} = \frac{\partial \varphi}{\partial x_{i'}'}\frac{\partial x_{i'}'}{\partial x_i} = F_{i'i}k'_{i'}.
\end{align}
Primed indices are associated with the transformed coordinates. Using Eqs.~\eqref{eq:ddx_transform} and \eqref{eq:k_transform} in Eq.~\eqref{eq:Bloch_EVP_periodic}, the Bloch eigenvalue problem for the reference unit cell rewritten in transformed coordinates is
\begin{equation}\label{eq:EVP_2}
\begin{aligned} 
    (C_{ijkl}F_{l'l}F_{j'j})_{,j'} (A_{k,l'}+\text{i}k'_{l'}A_k) + C_{ijkl} F_{l'l}F_{j'j}(A_{k,l'j'}+\text{i} k'_{j'}A_{k,l'} +\text{i} k'_{l'}A_{k,j'} &- A_k k'_{j'}k'_{l'}) \\
    &+ \rho \omega^2 \delta_{ik} A_k = 0,
\end{aligned}
\end{equation}
where $\delta_{ik}$ denotes the Kronecker delta. We now define the amplitude $\bfA'$ in the transformed coordinates such that $A_k=F_{k'k}A'_{k'}$. Multiplying Eq.~\eqref{eq:EVP_2} by $F_{i'i}$ yields
\begin{equation}\label{eq:EVP_3}
\begin{aligned} 
    &(C_{ijkl}F_{i'i}F_{j'j}F_{k'k}F_{l'l})_{,j'} (A'_{k',l'}+\text{i}k'_{l'}A'_{k'}) \\
    \ \ \ &+ C_{ijkl}F_{i'i}F_{j'j}F_{k'k}F_{l'l} (A'_{k',l'j'}+\text{i} k'_{j'}A'_{k',l'} +\text{i} k'_{l'}A'_{k',j'} - A'_{k'} k'_{j'}k'_{l'})
     + \rho \omega^2 \delta_{ik} F_{i'i} F_{k'k} A'_{k'} = 0.
\end{aligned}
\end{equation}
This can be simplified by assuming that each point within the unit cell has isotropic material properties. The corresponding elastic tensor takes the form $C_{ijkl}=\lambda\delta_{ij}\delta_{kl}+\mu(\delta_{ik}\delta_{jk}+\delta_{il}\delta_{jk})$, where $\lambda$ and $\mu$ are the first and second Lam\'e parameters, respectively. Using Eq.~\eqref{eq:jacobian} and the fact that rotation matrices are orthogonal ($R_{ij}R_{kj}=\delta_{ik}$), the terms in Eq.~\eqref{eq:EVP_3} simplify to
\begin{equation} \label{eq:EVP_simplifications1}
    C_{ijkl}F_{i'i}F_{j'j}F_{k'k}F_{l'l} = J^4 R_{i'i}R_{j'j}R_{k'k}R_{l'l}\left[\lambda\delta_{ij}\delta_{kl}+\mu(\delta_{ik}\delta_{jk}+\delta_{il}\delta_{jk})\right]
    =J^4 C_{i'j'k'l'}
\end{equation}
and
\begin{align} \label{eq:EVP_simplifications2}
      F_{i'i}F_{k'k}\delta_{ik} = J^2 R_{i'i}R_{k'k}\delta_{ik} = J^2\delta_{i'k'}.
\end{align}
Inserting Eqs.~\eqref{eq:EVP_simplifications1} and \eqref{eq:EVP_simplifications2} into Eq.~\eqref{eq:EVP_3}, we obtain
\begin{equation}
\begin{aligned}
    J^4 C_{i'j'k'l',j'}(A'_{k',l'}+\text{i}k'_{l'}A'_{k'}) + J^4 C_{i'j'k'l'}(A'_{k',l'j'}+\text{i} k'_{j'}A'_{k',l'} +\text{i} k'_{l'}A'_{k',j'} - &A'_{k'} k'_{j'}k'_{l'}) \\
    &+ J^2 \rho \omega^2 A'_{i'} = 0,
\end{aligned}
\end{equation}
which can be rewritten as
\begin{align} \label{eq:EVP_final}
    \boldsymbol{\mathcal{L}'}^B [ \bfA' ] + J^{-2}\omega^2 \rho \bfA' = \boldsymbol{0}.
\end{align}
Here, $\boldsymbol{\mathcal{L}'}^{B}$ represents the Bloch operator of Eq.~\eqref{eq:Bloch_operator} written in the transformed coordinates. Finally, upon defining $\omega'=J^{-1} \omega$, Eq.~\eqref{eq:EVP_final} recovers the Bloch eigenvalue problem in the transformed coordinates:
\begin{align} \label{eq:EVP_final}
    \boldsymbol{\mathcal{L}'}^B [ \bfA' ] + \omega' \rho \bfA' = \boldsymbol{0},
\end{align}
which is identical to what we would obtain if we directly wrote the eigenvalue problem for the scaled and rotated unit cell. Therefore, the dispersion relations $\omega'(\bfk')$ of a scaled and rotated unit cell are directly obtained from the dispersion relations $\omega(\bfk)$ of the reference unit cell by
\begin{align}
    \omega'(\bfk') = J^{-1} \omega(\bfk).
\end{align}
While this derivation holds strictly for periodic lattices, the results also apply to the local dispersion relations of locally periodic lattices under the assumption that $\bfF$ is (approximately) constant throughout each unit cell. This means that in a conformally graded metamaterial, assumed to exhibit geometrically similar unit cells, the dispersion relations of a single reference unit cell describes the wave propagation throughout the entire metamaterial. All local dispersion relations can be obtained directly from those of the reference unit cell.


\section{Planar truss lattice}

This section provides details about the geometry, experimental setup, and numerical modeling for the planar truss lattice of Figs.~1 and 2 of the main text.


\subsection{Conformal scaling factor}
From the complex representation of the conformal mapping $z'=\exp(z)$, Eq.~(2) of the main text, the scaling factor $J$ can be computed. The scaling factor of a conformal transformation in terms of complex function $z'(z)$ is given by \cite{nehari2012conformal}
\begin{align}
    J=\left|\frac{dz'}{dz}\right|.
\end{align}
Upon inserting Eq. (2) of the main text, this becomes
\begin{align} \label{eq:radial_scaling_factor}
J=\left|e^z\right|=\left|e^{\ln z'}\right| = r',
\end{align}
where $r'=\sqrt{(x'_1)^2+(x'_2)^2}$ is the radial coordinate in the transformed geometry. Thus, the scaling factor increases linearly with the radius.


\subsection{Experiment}

Here, we describe the details of the fabrication of the truss lattice prototype of Fig.~2a and of the experiment that produced the results of Fig.~2d in the main text.


\subsubsection{Fabrication}
\label{fabrication}

A prototype of the planar truss lattice was fabricated from an aluminum alloy 5083 plate with dimensions 63$\times$63$\times$0.4~cm. The prototype consists of a truss region that ranges from an inner radius of $r'_\text{i}=7.96$~cm to an outer radius of $r'_\text{o}=29.77$~cm. Within the truss region, there are 100 unit cells around the circumference and 21 unit cells along the radial direction. The regions inside $r'_\text{i}$ and outside $r'_\text{o}$ are solid material. Dimensions of the prototype were scaled such that the innermost unit cell (at $r'=r'_\text{i}$) has a width of 0.5~cm and a beam width of 0.5~mm. 

Since the unit cell scaling factor varies linearly with radius, as evident from Eq.~\eqref{eq:radial_scaling_factor}, the thickness of the prototype was linearly tapered to ensure that all unit cells are geometrically similar. The outer region of the prototype ($r'>r'_\text{o}$) has a thickness of 4~mm, resulting in a thickness of 1.07~mm in the inner region ($r'<r_\text{i}$). Thus, the lattice region between $r'_\text{o}$ and $r'_\text{i}$ exhibits a linear variation in thickness along the radial coordinate from 1.07 to 4~mm, which can be seen in the close-up photo of the prototype in Fig.~2b of the main text.

Aluminum was chosen for the material because it exhibits low damping. This is beneficial for our experimental demonstration to ensure that the observed wave attenuation is primarily due to the geometric structure of the lattice as opposed to material loss. 

Fabrication of the prototype lattice was performed as a two-step process. First, the radial thickness taper between $r'_\text{i}$ and $r'_\text{o}$ was cut into the plate by CNC milling, resulting in a solid plate with a thickness taper on one side and a flat surface on the other side. Then, the truss section between $r'_\text{i}$ and $r'_\text{o}$ was cut by water jet to complete the fabrication process.


\subsubsection{Measurement and data processing}
\label{sec:measurement_data_processing}
\begin{figure}
    \centering
    \includegraphics[width=\linewidth]{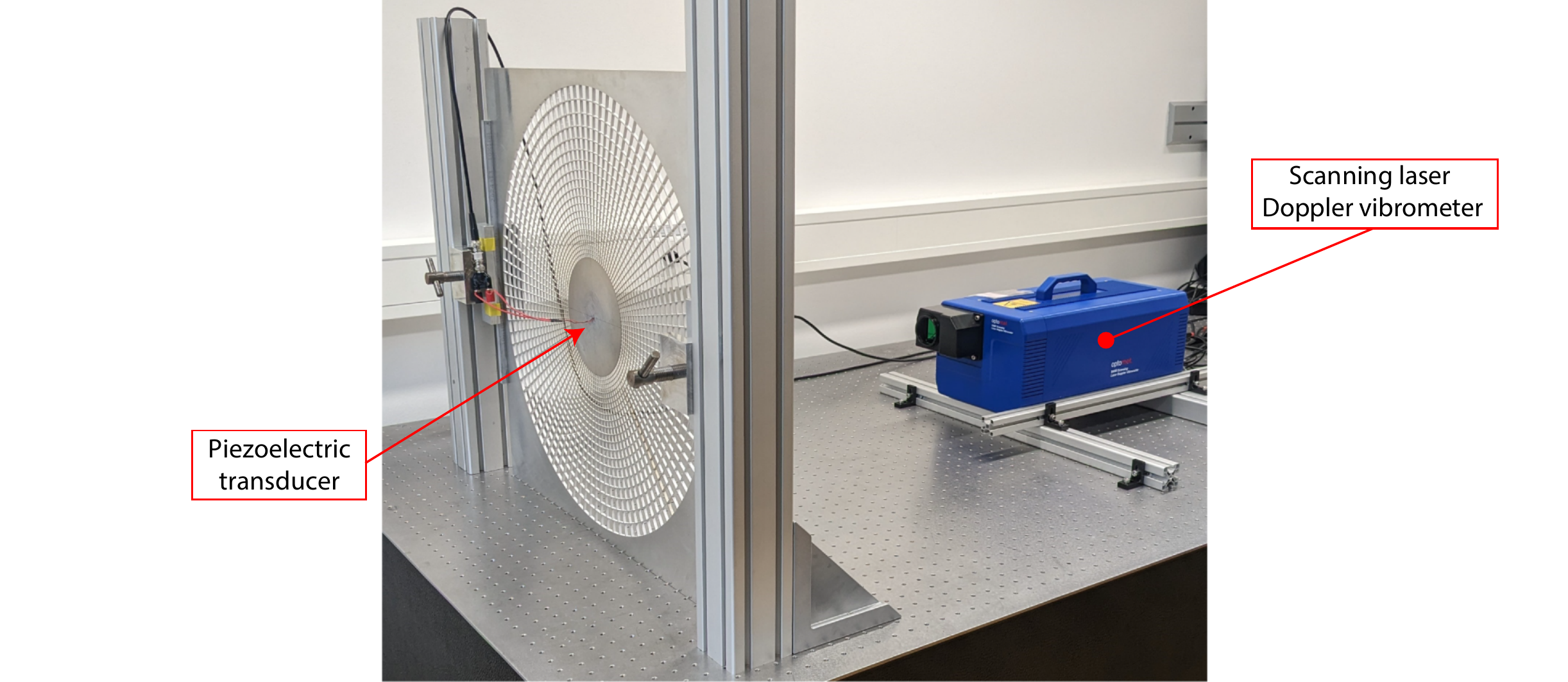}
    \caption{Experimental setup, consisting of a piezoelectric transducer that excites out-of-plane waves in the prototype, whose response is measured by a laser Doppler vibrometer.}
    \label{fig:figureS2}
\end{figure}

The experimental setup for measuring wave propagation in the prototype is shown in Fig.~\ref{fig:figureS2} and illustrated in Fig.~2c of the main text. To excite out-of-plane elastic waves, a piezoelectric transducer (Thorlabs PA3BCW) was glued to the center of the prototype. A function generator (Teledyne LeCroy T3AFG40) was used to generate an input signal, which was sent through an amplifier (Thorlabs HVA200) to the transducer. The input signal was a swept sine ranging from 50 to 300~kHz with a linear sweep rate and a duration of 100~$\mu$s. The swept sine was shifted to have a minimum value of zero to avoid applying negative voltage to the transducer. This input signal was observed to provide sufficient broadband excitation in the 0$-$300~kHz range in the truss region.

A scanning laser Doppler vibrometer (Optomet SWIR) was used to measure the response of the prototype. Measurements were collected at 100 evenly spaced points along radial lines from $r'_\text{i}$ to $r'_\text{o}$. At each measurement point, the input signal was repeatedly applied 1~second apart to allow the vibrations to decay before the next measurement. For each measurement, the displacement signal was recorded with a sampling rate of 4~MHz over a duration of 16.384~ms. To reduce noise, 100~independent measurements were averaged at each measurement point. The maximum observed displacement throughout all measurements was 18.0~nm. No significant differences in the results were observed along different radial lines.

To quantify the wave attenuation along a radial line, we estimate the frequency response function (FRF) between location $r'$ and the inner radius $r'_\text{i}$. We interpret the displacement signal at $r'_\text{i}$ as an input to the truss region, and we observe the attenuation of this input using the FRF. To estimate the FRF, following \cite{ewins2009modal}, the power spectral density of the displacement signal at $r'_\text{i}$ is denoted $S_{r'_\text{i} r'_\text{i}}$ and the cross power spectral density between the response at $r'_\text{i}$ and $r'$ is denoted $S_{r'_\text{i} r'}$. The spectral densities are computed using overlap averaging (Welch's) method with temporal segments of 2~ms, an overlap of 1~ms, and a Tukey window applied to each segment (with the flat region making up 75\% of the window). The FRF, denoted $H(r',f)$ is estimated from the spectral densities by
\begin{align}
H(r',f) = \frac{S_{r'_\text{i} r'}}{S_{r'_\text{i} r'_\text{i}}}.
\end{align}
In Fig.~2d of the main text, the amplitude of the FRF is plotted in decibels, which is $20 \log_{10} |H(r',f)|$.

\subsection{Numerical modeling}
To model the truss lattice numerically, we perform a dispersion relation scaling analysis (Section~\ref{sec:dispersion_scaling}) based on numerically computed dispersion relations to predict at which radius each frequency becomes forbidden. This scaling analysis is validated by comparing it to a transient finite element simulation of the entire truss lattice (Section~\ref{sec:transient_FE}). Finally, we discuss the use of higher-fidelity modeling using 3D finite elements instead of beam elements (Section~\ref{sec:3D_modeling}). All finite element computations are performed using the state-of-the-art finite element code ae108 \cite{ae108}.


\subsubsection{Dispersion relation scaling analysis} \label{sec:dispersion_scaling}

This section describes the dispersion relation scaling analysis that produces Fig.~2e of the main text. The aim is to compute how far a wave of each frequency starting at the inner radius can propagate radially, before the unit cell scaling forbids propagation at that frequency. 

We begin by numerically computing the dispersion relations of the innermost unit cell using a finite element model, approximating the unit cell as a square. To match the experimental dimensions (see Section~\ref{fabrication}), we take a unit cell with an edge length of $l=0.5$~cm, a thickness of 1.07~mm, and 0.5~mm wide beams along its edges. The material is modeled as homogeneous, isotropic linear elastic with the following properties to model aluminum: Young's modulus $E=69.9$~GPa, Poisson's ratio $\nu=0.33$, and mass density $\rho=2700$~kg/m$^3$. Each beam is discretized by 40 Timoshenko beam finite elements.

\begin{figure}
    \centering
    \includegraphics[width=\linewidth]{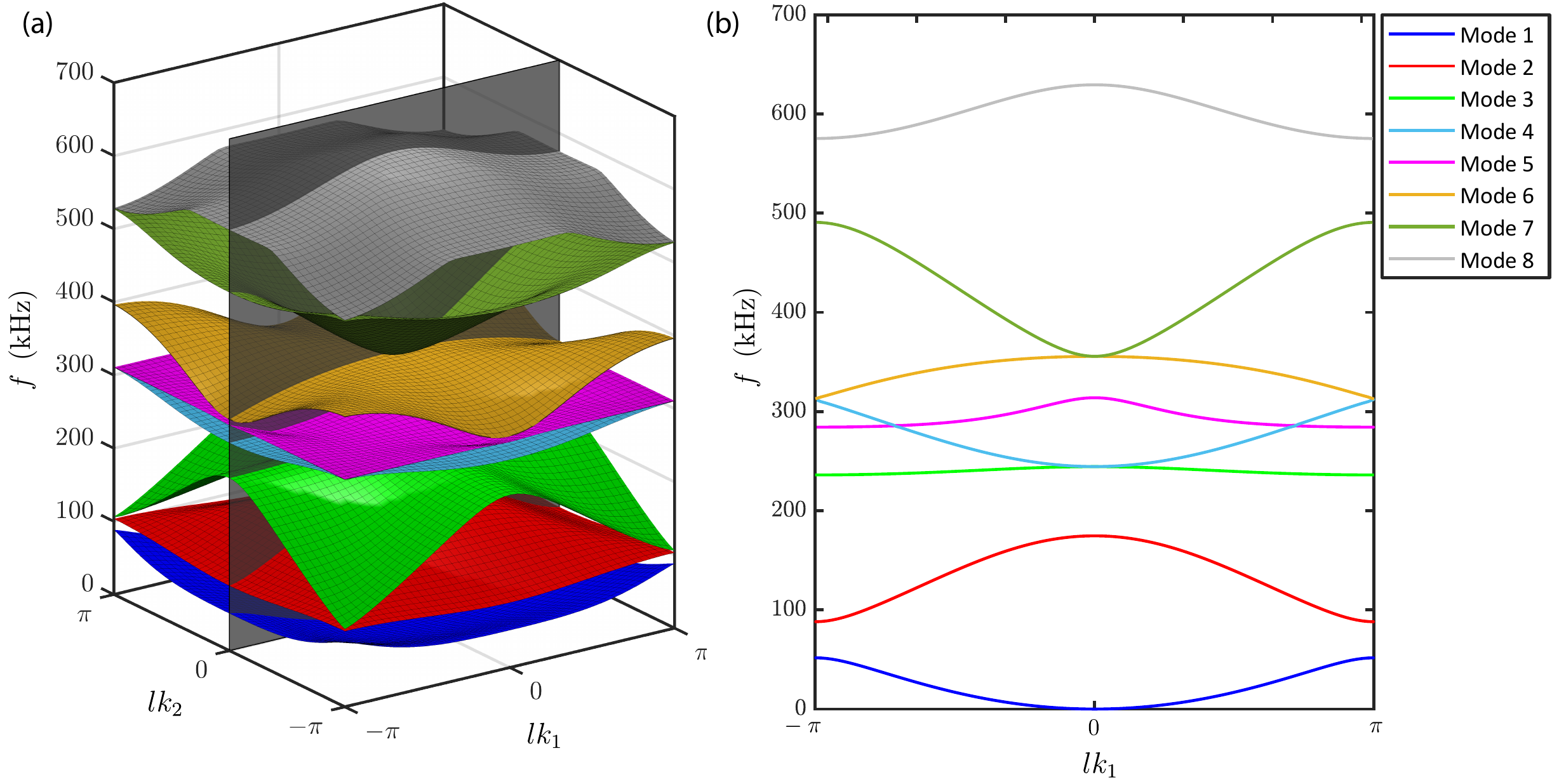}
    \caption{(a) Dispersion relations of the innermost unit cell of the radial lattice, with the $k_2=0$ plane shaded gray. (b) Slice of the dispersion relations intersecting $k_2=0$.}
    \label{fig:figureS3}
\end{figure}

To compute the dispersion relations from the finite element model of the unit cell, we follow the formulation presented in~\cite{zelhofer2017acoustic}. For simplicity, we consider only out-of-plane displacements since in- and out-of-plane modes are decoupled for planar truss lattices \cite{zelhofer2017acoustic}. That is, in response to out-of-plane excitation we can expect predominately out-of-plane displacements. The first eight out-of-plane dispersion surfaces are plotted in Fig.~\ref{fig:figureS3}a.

To proceed with the dispersion relation scaling analysis, we first note that the unit cells in the truss lattice do not vary circumferentially. Thus, the local dispersion relations are axisymmetric. We also consider a point excitation at the origin, so that the loading is also axisymmetric. As a result, waves can only propagate radially. Since the unit cell is aligned with the radial and circumerential coordinates (we take $k_1$ to align with the radial coordinate), $k_2=0$ due to axisymmetry. The $k_2=0$ plane is shaded in gray in Fig.~\ref{fig:figureS3}a. The corresponding slice of the dispersion relations in the $k_2=0$ plane is plotted in Fig.~\ref{fig:figureS3}b.

With the dispersion curves in the $k_2=0$ plane in hand, we perform a scaling analysis. Consider a specific frequency $f_r$ that intersects mode 2 in Fig.~\ref{fig:figureS3}b. The maximum frequency of mode 2's dispersion curve is $f_{2,max}=174.37$~kHz. Therefore, if the unit cell is scaled up by a factor of $J>f_{2,max}/f_r$, mode~2's dispersion curve will scale down and no longer intersect $f_r$ and this frequency is forbidden. Using Eq.~\eqref{eq:radial_scaling_factor}, the critical frequency is converted to a critical radius, beyond which propagation is forbidden.

A similar analysis is performed for mode~1, except that frequencies below some low-pass cutoff $f_c$ will reach the outer radius. The outermost unit cell is $J_\text{o}=3.51$ times larger than the innermost one (this value is taken to match the prototype). Since the maximum frequency of mode~1 in Fig.~\ref{fig:figureS3}b is $f_{1,max}=51.66$~kHz, it follows that frequencies below the cutoff $f_c=f_{1,max}/J_\text{o}=14.72$~kHz have a critical scaling factor larger than $J_\text{o}$ and will reach the outer radius.

Extra care must be taken in the scaling analysis of modes 3 to 7, because there are degeneracies (repeated eigenfrequencies for a given wave vector) where dispersion surfaces intersect. Property (ii) of the main text is violated for degenerate frequencies, and we assume that mode conversion can happen between modes 3-7. Instead of scaling each individual mode, we scale the block of modes 3 to 7, where sufficient scaling is required such that $f_r$ does not intersect the range of frequencies spanned by modes 3 to 7. Following this procedure, we obtain the uppermost region of allowed propagation in Fig.~2e of the main text.


\subsubsection{Transient finite element simulation}
\label{sec:transient_FE}
For validation of the dispersion scaling analysis, we performed a transient finite element simulation of the truss lattice. This simulation provides validation of the assumptions of the dispersion relation scaling analysis (local periodicity and idealization of unit cells as squares) in a controlled and noise-free setting to complement the experimental analysis.

The truss lattice geometry (see Fig.~1b of the main text) was discretized with Timoshenko beam elements. The out-of-plane displacement of all nodes on the inner radius was prescribed to provide broadband excitation. Specifically, a swept sine ranging from 50 to 300~kHz with a linear sweep rate and a duration of 100~$\mu$s was applied. Using a Newmark-beta solver for implicit time integration, the solution was computed for a duration of 800~$\mu$s with timesteps of 0.4~$\mu s$.

\begin{figure}
    \centering
    \includegraphics[width=\linewidth]{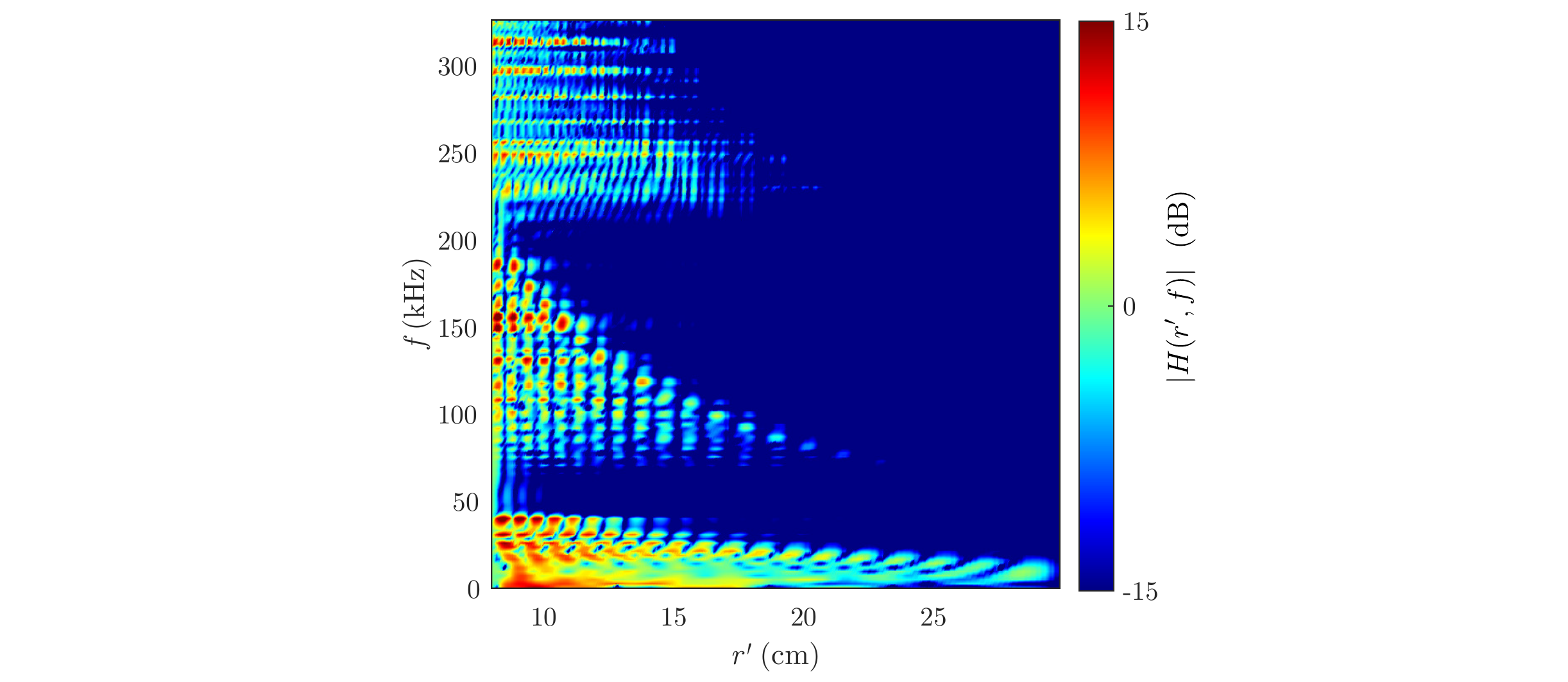}
    \caption{Transient finite element simulation results for the planar truss lattice. The FRF magnitude along one radial line is plotted.}
    \label{fig:figureS4}
\end{figure}

Displacement signals along one radial line were processed using the same procedure as for the experiment (Section \ref{sec:measurement_data_processing}) to compute the FRF with respect to the displacement applied to the inner radius. Fig.~\ref{fig:figureS4} shows the FRF amplitude as a function of frequency and radius. The result is comparable to the analogous experimental result shown in Fig.~2d of the main text. However, cleaner results are available here, since there is no noise, perfect symmetry, and an ideal broadband excitation was applied directly to the inner radius of the truss. Results show close agreement between the dispersion scaling analysis (Fig.~2e of the main text) and Fig.~\ref{fig:figureS4}. Note that in the transient finite element simulation, some frequencies propagate slightly further than predicted by the dispersion scaling analysis. This is due to the fact that the wave is not immediately attenuated when it reaches a forbidden region; instead, this is where the wave begins to exponentially decay.

\subsubsection{Comparison to higher fidelity modeling}
\label{sec:3D_modeling}

The above numerical analysis relied on modeling truss lattices as networks of beams. The chosen beam model captured the behavior observed in the experiment while also providing a computationally efficient model. However, higher-fidelity modeling is possible using 3D finite elements at the cost of computational efficiency. We show here that there are differences in the dispersion relations computed by solid 3D elements versus beam finite elements \cite{TelgenThesis2022}, but these differences have a minimal effect on the wave attenuation behavior of the planar truss lattice.

\begin{figure}
    \centering
    \includegraphics[width=\linewidth]{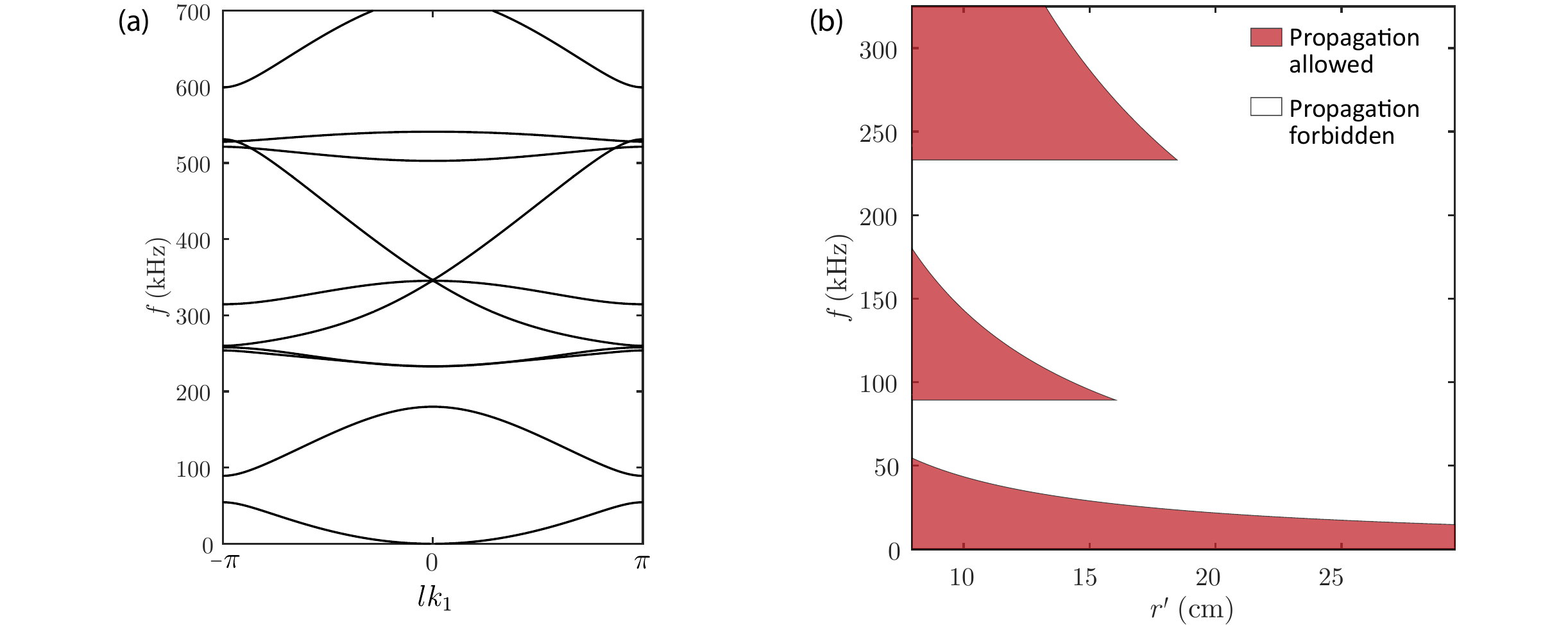}
    \caption{(a) Dispersion relations for the same unit cell as in Fig.~\ref{fig:figureS3} computed using a 3D finite element model along the $k_2=0$ plane. (b)~Regions of forbidden (white) and allowed (red) linear elastic wave propagation, computed based on the dispersion relations of (a).}
    \label{fig:figureS5}
\end{figure}

To compute the dispersion relations of a square unit cell (with the same geometry and material properties as used in Section \ref{sec:dispersion_scaling}) using 3D finite elements, the unit cell is discretized by tetrahedral elements, using 7 elements through the width and 14 through the thickness of each beam---following a mesh convergence study, resulting in a mesh with 61{,}923 elements in total. Bloch boundary conditions are applied, and the Bloch eigenvalue problem is solved for points from  $k_1=0$ to $k_1=\pi l$ (where $l$ is the unit cell edge length) with $k_2=0$. For each value of $\bfk$, the computation time was approximately 10 minutes (in contrast to a 0.1 second computation time for each $\bfk$-point when using the beam model). The resulting dispersion relations are plotted in Fig.~\ref{fig:figureS5}a. Here, only modes with mode shapes exhibiting a nonzero out-of-plane component are shown; all purely in-plane modes are omitted (the validity of this assumption and the primary relevance of the out-of-plane modes is confirmed by the experimental data).

The lowest two dispersion curves of Fig.~\ref{fig:figureS5}a closely match those of the beam model (Fig.~\ref{fig:figureS3}b). However, there are differences in the higher dispersion curves between the 3D and beam models. The prediction for the regions of allowed and forbidden propagation based on the 3D model's dispersion relations are plotted in Fig.~\ref{fig:figureS5}b. The partial bandgap between 500 and 600~kHz is present in both models, which results in nearly identical predictions (Figs.~\ref{fig:figureS5}b and 2d of the main text) for the regions of allowed and forbidden propagation. Due to the fact that these predictions are in agreement between the beam and 3D models, we focus on the beam models for computational efficiency, though it is important to address the availability of higher-fidelity modeling, which may become important in other scenarios.


\section{Curved metamaterial surfaces}

This section provides details about the design and simulation of the curved metamaterial surfaces presented in Figs.~3 and~4 of the main text.

\subsection{Discrete conformal equivalence}

Our objective is to transform the periodic reference lattice into a surface that realizes a prescribed conformal scaling factor distribution $J(\bfx)$. This is achieved by re-scaling the edge lengths $l$ of the reference lattice to spatially varying target lengths $\tilde{l}$. To ensure that the transformation is conformal, the notion of conformality must be defined in the discrete setting. Allowing each edge to scale independently is not sufficiently restrictive, since it allows for arbitrary scaling. Alternatively, since conformal mappings are angle preserving in the continuous setting, one could enforce angle preservation in the discrete setting, but this is too restrictive since it allows only global scaling and rotation \cite{Crane:2020:DCG}. An appropriate definition of discrete conformality is given by discrete conformal equivalence \cite{springborn2008conformal}, which preserves the basic structure of continuous conformal geometry in the discrete setting. 

Consider an edge $ij$ of the reference lattice, which connects vertices $i$ and $j$ and has length $l_{ij}$. The transformed lattice with re-scaled edge lengths $\tilde{l}_{ij}$ is \textit{conformally equivalent} to the reference lattice if all edge lengths satisfy
\begin{align} \label{eq:discrete_conformal_equivalence}
    \tilde{l}_{ij} = e^{\frac{1}{2}(U_i+U_j)}l_{ij},
\end{align}
where $U$ is the \textit{log-scaling factor}, which is defined on the vertices ($U_i$ and $U_j$ being the log-scaling factors of vertices $i$ and $j$). 

Thus, we prescribe the target log-scaling factor $U$ on each vertex, from which the set of target edge lengths $\tilde{l}$ that ensure conformality follow from Eq. \eqref{eq:discrete_conformal_equivalence}. The resulting conformal scaling factor associated with each edge is $J_{ij} = e^{\frac{1}{2}(U_i+U_j)}$.


\subsection{Conformal surface optimization}

In the conformal bump example of Fig.~3 in the main text, the log-scaling factor was prescribed according to a radial bump function of the form
\begin{align}
    U(r)=
    \begin{cases}
    3 \exp \left(\frac{-1}{(1-\frac{r}{20l})^2}\right) & r<20l
    \\
    0 & r\geq20l.
    \end{cases}
\end{align}
Here, $r=\sqrt{x_1^2+x_2^2}$ is the radial coordinate in the reference domain and $l$ is the side length of the reference unit cell. The log-scaling factor is evaluated at each vertex location, i.e., $U_i=U(\bfx_i)$. Then, the target edge lengths $\tilde{l}$ are computed according to Eq.~\eqref{eq:discrete_conformal_equivalence}.

In the conformal spiral example of Fig.~4 in the main text, the log-scaling factor was prescribed along an Archimedean spiral of the form $r=5.75l\theta$. The reference unit cell has side length $l$, while $r$ and $\theta$ are polar coordinates in the reference domain. The spiral was given a thickness of 5 unit cells; vertices inside the spiral were prescribed $U=-0.75$ while unit cells outside were prescribed $U=0$. Finally, a Gaussian filter with a standard deviation of 1.5$l$ was applied to spatially smooth the distribution of $U$ on the reference lattice, thus removing sharp jumps in the target scaling factor.

The definition of conformal equivalence from Eq.~\eqref{eq:discrete_conformal_equivalence} of the main text holds for triangulated discrete surfaces \cite{springborn2008conformal}. The lattices considered, however, contain quadrilateral unit cells. To ensure that conformal equivalence is satisfied, we triangulated the lattice during the optimization by adding one diagonal edge to triangulate each unit cell. This edge is only present during the optimization; once the optimal geometry is obtained, the triangulated edge is removed.

For both examples, Matlab's \textit{fminunc} utility was used to perform the optimization using an interior point algorithm with cost function gradients provided, which were derived analytically.  When performing the optimization, the ratio of the cost function weights in Eq.~(4) of the main text was taken as $\kappa/k=5\times10^{-5}$ for the conformal bump example, and $\kappa/k=7.5\times10^{-5}$ for the conformal spiral example. For the conformal bump example, the constraints of fixing the boundaries of the surface were directly applied by eliminating the boundary vertex positions from the design variables.


\subsection{Reference dispersion relations}
\label{sec:dispersion_relations_surface}

\begin{figure}
    \centering
    \includegraphics[width=\linewidth]{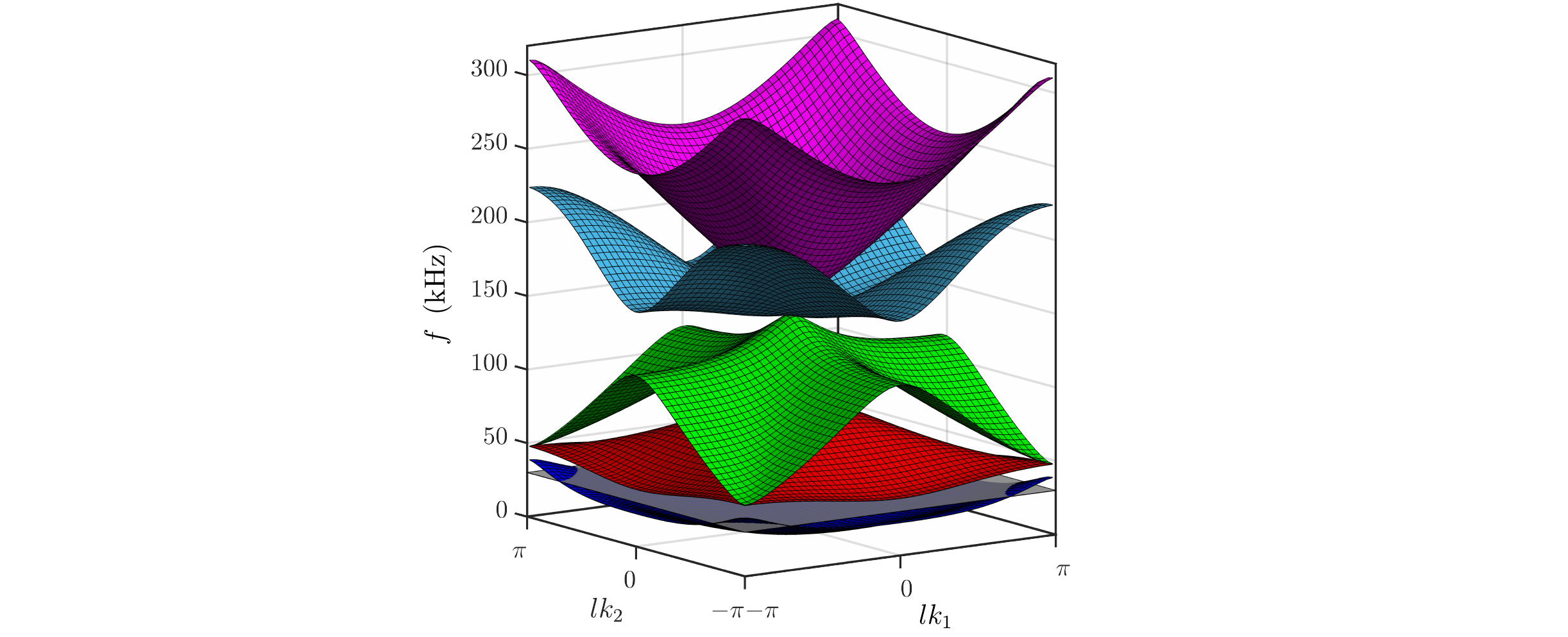}
    \caption{Dispersion relations at the excitation location of the two curved surface examples. The plane corresponding to $\omega=30$~kHz is highlighted in gray.}
    \label{fig:figureS6}
\end{figure}

For both examples, the beams that comprise the truss lattice are given a circular cross-section, which is more straightforward to define on a curved lattice than the previously considered rectangular cross-sections. To compute the dispersion relations of a reference unit cell, we take the unit cell size as 0.5$\times$0.5~cm with the same material properties as in the planar truss lattice example. The cross-section's diameter is taken as $0.5$~mm. The dispersion relations, computed by the same process as outlined in Section~\ref{sec:dispersion_scaling} but with a circular cross section, are plotted in Fig.~\ref{fig:figureS6}.

\subsection{Transient finite element simulations}

Transient finite element simulations were performed to demonstrate the wave attenuation capabilities of the curved metamaterial surface examples.

The unit cell edge length at the excitation location was set to 0.5~cm for both the conformal bump and spiral examples, so that the local dispersion relations at the excitation location are given in Fig.~\ref{fig:figureS6}. Throughout the lattice, the cross section of each beam is circular with a diameter of $1/10$ of its length to ensure that all unit cells are approximately geometrically similar. Again, the material properties of aluminum were used, identical to those used in the example of Section~\ref{sec:dispersion_scaling}. Timoshenko beam finite elements were used to discretize the lattice.

For the conformal bump surface example, displacement excitation was applied to one boundary of the surface (highlighted in red in Fig.~3b of the main text) in the $\bfx'_3$-direction, which is locally out-of-plane. The displacement excitation is a swept sine ranging from 50 to 300~kHz with a linear sweep rate; Fig.~\ref{fig:figureS6} highlights the modes present in this frequency range.

For the conformal spiral surface example, a harmonic point excitation with a frequency of 30~kHz was applied at the center of the spiral in the locally out-of-plane direction, as shown in Fig.~4b of the main text. Based on the dispersion relations of Fig.~\ref{fig:figureS6}, this frequency only excites mode 1, but it is near the maximum frequency of the first dispersion surface such that a small increase in unit cell scaling will forbid propagation.

In both examples, a Newmark-beta time integration scheme was used. For the conformal bump example, the simulation was performed over a duration of 1.2~ms with a fixed time step of 2.4~MHz. For the conformal spiral example, the simulation was performed over a duration of 2.5~ms with a fixed time step of 500~kHz. The maximum value of the norm of the three-dimensional displacement vector during the simulation is plotted in Figs.~3b and 4b of the main text. Animations of the simulations are included in Supplementary Videos S2 and S3.


\section{Supplementary video legends}

\textbf{Video S1}: Animation of how a dispersion surface scales with unit cell size. The second out-of-plane dispersion surface for a square unit cell is shown, whose frequency scales inversely with the unit cell edge length.

\textbf{Video S2}: Finite element simulation of the conformal bump. Broadband excitation is applied to the left edge (red) and the resulting displacement magnitudes are shown. Large unit cells on the bump are isolated from the incident waves.

\textbf{Video S3}: Finite element simulation of the conformal spiral. Harmonic excitation is applied locally out-of-plane at the spiral center and the resulting displacement magnitudes are shown. The wave is follows the path of small unit cells and is guided along the spiral.


\providecommand{\noopsort}[1]{}\providecommand{\singleletter}[1]{#1}%

\end{document}